\documentclass[aps,prl,preprintnumbers,notitlepage,tightenlines,nofootinbib,twocolumn,superscriptaddress]{revtex4-2}

\usepackage{multirow}

\usepackage{amsmath}
\usepackage{amssymb,amsthm}
\usepackage{mathtools}
\usepackage{mathrsfs}
\usepackage{relsize}
\usepackage{bm}
\allowdisplaybreaks[4]  
\usepackage{epsfig,graphics,graphicx}
\usepackage{color,xcolor}
\usepackage{subfigure}
\usepackage{array}
\usepackage{ragged2e}
\usepackage{lineno}
\usepackage{natbib}
\usepackage{hyperref}
\hypersetup{colorlinks=true,
            linkcolor=blue,
            anchorcolor=blue,
            citecolor=magenta,
            filecolor=blue,
            urlcolor=blue,
            bookmarksnumbered=true,
            pdfview=FitB
}
\usepackage{enumitem}
\usepackage{ulem}
\usepackage{listings}
\usepackage{soul}
\usepackage{cancel}

\colorlet{darkgreen}{green!50!black}
\colorlet{brightyellow}{yellow!75!red}
\colorlet{orange}{red!50!yellow}
\colorlet{darkgray}{gray!50!black}
\colorlet{darkred}{red!50!black}

\definecolor{XHCmathematicaTitleRed}{RGB}{204,11,2}

\def\dd{{\mathrm{d}}}

\newcommand{\half}[1][1] {\mathsmaller{\frac{#1}{2}}}

\makeatletter
\newcommand*{\transpose}{%
  {\mathpalette\@transpose{}}%
}
\newcommand*{\@transpose}[2]{%
  \raisebox{\depth}{$\m@th#1\intercal$}%
}
\makeatother

\begin{document}

\title{Origin of the nucleon gravitational form factor $B_N(t)$: exposition in light-front holographic QCD}

\author{Xianghui Cao}
\affiliation{Department of Modern Physics, University of Science and Technology of China, Hefei 230026, China}

\author{Bheemsehan Gurjar}
\affiliation{Department of Modern Physics, University of Science and Technology of China, Hefei 230026, China}
\affiliation{Department of Physics, Indian Institute of Technology Kanpur, Kanpur-208016, India}

\author{Chandan Mondal}
\affiliation{Institute of Modern Physics, Chinese Academy of Sciences, Lanzhou 730000, China}
\affiliation{School of Nuclear Science and Technology, University of Chinese Academy of Sciences, Beijing 100049, China}

\author{Chen Chen}
\affiliation{Interdisciplinary Center for Theoretical Study, University of Science and Technology of China, Hefei, Anhui 230026, China}
\affiliation{Peng Huanwu Center for Fundamental Theory, Hefei, Anhui 230026, China}

\author{Yang Li}
\thanks{Corresponding author}
\affiliation{Department of Modern Physics, University of Science and Technology of China, Hefei 230026, China}
\affiliation{Anhui Center for Fundamental Sciences in Theoretical Physics, Hefei, 230026, China}

\preprint{USTC-ICTS/PCFT-26-05}

\date{\today}

\begin{abstract}
Recent lattice QCD simulations and phenomenological models indicate that the nucleon's gravitational form factor $B_N(t)$ remains remarkably small at finite momentum transfer $t$. While $B_N(0) = 0$ is a known consequence of the equivalence principle, the physical origin of its suppression at finite $t$ has not been fully elucidated. In this work, we demonstrate that the smallness of $B_N(t)$ arises from a fundamental cancellation within the nucleon's wave functions. Using light-front holographic QCD, we show that $B_N(t)$ is governed by an antisymmetric factor in the longitudinal dynamics that leads to the exact vanishing of the form factor in the symmetric limit and significant suppression for realistic nucleon structures. Our results suggest that the smallness of $B_N(t)$ is a signature of the nucleon's dominant S-wave character, providing a formal justification for its frequent omission in practical applications like near-threshold $J/\psi$ production.

\end{abstract}

\maketitle

Gravitational form factors (GFFs) characterize the distribution of energy, spin, and pressure within the nucleon. They are defined via the hadron matrix element of the energy-momentum tensor, $T^{\mu\nu}$  \cite{Kobzarev:1962wt, Pagels:1966zza, Ji:1996ek}:
\begin{multline}
\langle p', \sigma'|T^{\mu\nu}(0)|p, \sigma\rangle =
\bar{u}_{\sigma'}(p')\Big[ A(t) \gamma^{\{\mu }P^{\nu\}} \\
+ B(t) \frac{iP^{\{\mu}\sigma^{\nu\}\rho}q_\rho}{2M}
+ D(t) \frac{q^\mu q^\nu - g^{\mu\nu}q^2}{4M}\Big] u_\sigma(p),
\end{multline}
where $P=(p+p')/2$, $q = p' - p$, and $t = q^2$, $a^{\{\mu}b^{\nu\}} = (a^\mu b^\nu + a^\nu b^\mu)/2$. Here, $A(t)$, $B(t)$, and $D(t)$ represent the three fundamental GFFs of the nucleon. Often, the form factor $J(t) = [A(t) + B(t)]/2$ is adopted in the literature. Consequently, investigating their properties is essential to unraveling the internal structure of the primary building blocks of matter.

\begin{figure}
\centering
\includegraphics[width=0.45\textwidth]{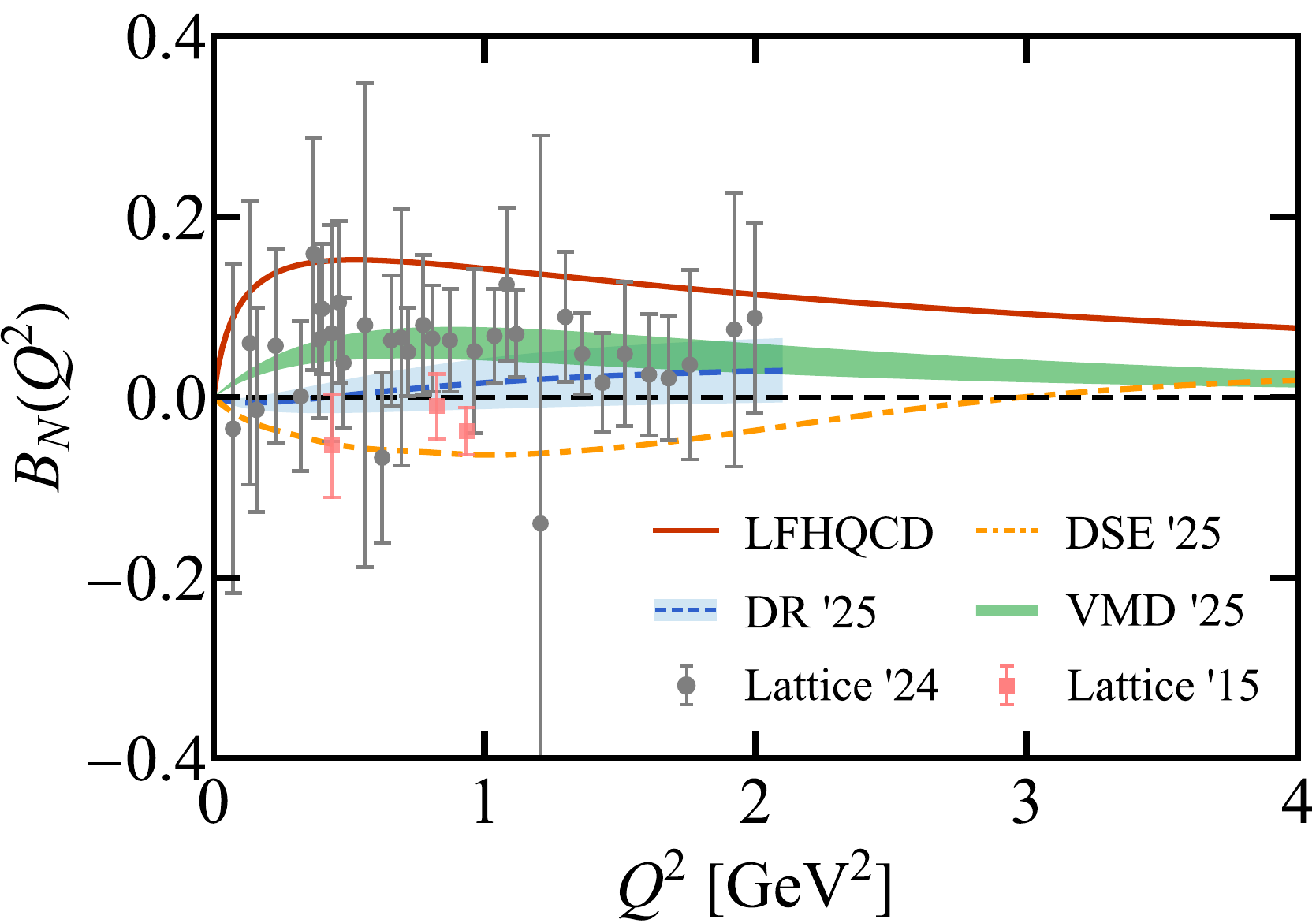}
\caption{
The gravitational form factor $B_N(t)$ of the nucleon obtained from lattice QCD (dots, \cite{Hackett:2023rif}),  dispersion relation and VMD analyses based on the lattice data (bands, \cite{Cao:2024zlf, Broniowski:2025ctl, Cao:2025dkv}), as well as two theoretical models, LFHQCD (solid line, see texts) and DSEs (dot-dash lines, \cite{Yao:2024ixu}). }
\label{fig:GFF_B}
\end{figure}

While the existing literature focuses predominantly on the GFFs $A(t)$ and $D(t)$, discussions regarding $B(t)$ remain relatively scarce \cite{Teryaev:1999su, Polyakov:2018zvc, Burkert:2023wzr}. This observable is intimately linked to the nucleon's spin structure; specifically, $B(0)$ corresponds to the anomalous gravitomagnetic moment (AGM). The equivalence principle (EP) requires that $B(0) = 0$ \cite{Teryaev:1999su}, a result proven in light-front QCD \cite{Brodsky:2000ii} and supported by lattice simulations \cite{Deka:2013zha, Hackett:2023rif} and model-based computations \cite{Cebulla:2007ei, Goeke:2007fp, Goeke:2007fq, Neubelt:2019sou, Azizi:2019ytx, Anikin:2019kwi, Chakrabarti:2020kdc, Fujita:2022jus, Nair:2024fit, Yao:2024ixu, Nair:2025sfr}. The vanishing of $B(0)$ is consistent with the tensor meson dominance in $\pi N$ scattering \cite{Renner:1970sbf, Jones:1971am}.
Furthermore, recent lattice QCD simulations suggest that the nucleon GFF $B_N(t)$ remains consistent with zero even at finite $t$, within current simulation uncertainties \cite{Deka:2013zha, Hackett:2023rif}. Although some dispersion relation analyses based on lattice data suggest a non-vanishing $B_N(t)$ as shown in Fig.~\ref{fig:GFF_B} \cite{Cao:2024zlf, Broniowski:2025ctl, Cao:2025dkv}, the extracted values are sufficiently small that they are often neglected in practical applications, such as predicting near-threshold $J/\psi$ photoproduction \cite{Hatta:2018ina, Mamo:2019mka, Hatta:2019lxo, Guo:2021ibg, Mamo:2021krl, Duran:2022xag, Guo:2024wxy, Guo:2025jiz, Meziani:2025dwu, Guo:2025muf}.
Regardless, the persistent smallness of $B_N(t)$ across different scales warrants a formal explanation.

One plausible explanation for the smallness of $B_N(t)$ is flavor cancellation \cite{Teryaev:1998iw}. Within the framework of generalized parton distributions (GPDs), the GFF $B_N(t)$ is identified as the second Mellin moment of the GPD $E(x, \xi, t)$:
\begin{equation}
B_{q,g}(t) = \int \dd x \, xE_{q,g}(x, \xi = 0, t),
\end{equation}
where $x$ is the longitudinal momentum fraction of the struck parton and $\xi$ is the skewness variable. In the proton, the GPDs $E_u$ and $E_d$ possess opposite signs. This is evident from their first Mellin moments at the forward limit ($t=0$), which correspond to the flavor-contributions to the anomalous magnetic moment \cite{Ossmann:2004bp}:
\begin{equation}
\begin{split}
\kappa_u =\,&  \int \dd x E_u(x, 0, 0)  \approx +1.67, \\
\kappa_d =\,& \int \dd x E_d(x, 0, 0)  \approx -2.03.
\end{split}
\end{equation}
The numerical values above are representative estimates from the constituent quark model. Unlike the Pauli form factor $F_2(t)$, which is defined by the first Mellin moment of $E$ weighted by the electric charges of the quarks, the GFF $B_N(t)$ is weighted by the momentum fraction $x$. Because $E_u$ and $E_d$ maintain their opposite signs across the $x$-distribution, a significant cancellation occurs between the $u$ and $d$ quark contributions. A simple non-relativistic quark model estimate yields:
\begin{equation}
\begin{split}
B_N(0) \approx B_u(0) + B_d(0)
\approx \,& \frac{2}{3} \kappa_u + \frac{1}{3} \kappa_d
\approx  0.44. \\
\end{split}
\end{equation}
While this simple estimate suggests a non-vanishing value, the result can be further refined. For instance, the MIT Bag Model predicts a vanishing AGM ($B_N(0) = 0$) in the static limit \cite{Ji:1997gm}. While relativistic corrections generally reintroduce a non-vanishing AGM, it has been shown that $B_N(t)$ vanishes in the large-$N_c$ limit \cite{Neubelt:2019sou}.
Finally, a complete description must account for the gluon contribution. At $t=0$, $B_g(0)$ is conjectured to vanish based on an extended EP \cite{Teryaev:1998iw, Teryaev:1999su}.

The suppression of $B_N(t)$ can be further understood through the fundamental symmetries governing graviton interactions. Within the Dyson-Schwinger equations (DSEs) framework, $B_N(t)$ is found to be vanishingly small, strictly obeying the constraint $B_N(0) = 0$ \cite{Yao:2024ixu}. This result is primarily driven by the symmetry of the graviton-quark vertex, which is rigorously enforced by Ward-Takahashi identities derived from the EP \cite{Brout:1966oea}. Complementary insights are provided by holographic QCD, which models hadron-graviton interactions within a 5D warped geometry \cite{Polchinski:2001tt, Abidin:2009hr, Hatta:2018ina, Mamo:2019mka, Fujita:2022jus, Fujii:2024rqd, Mamedov:2024tth, Deng:2025fpq, Liu:2025vfe, Nascimento:2025ryr, Fujii:2025eug, Nascimento:2025smf, Liu:2025qja, Hippelainen:2026izm}. In this framework, GFFs are computed via the scattering of a bulk graviton off a nucleon field in Anti-de Sitter (AdS) space. Notably, in models employing minimal gravitational coupling, $B_N(t)$ vanishes identically across all momentum transfers \cite{Abidin:2009hr}.
This result is fundamentally rooted in the fact that the minimal interaction vertex in the 5D bulk conserves helicity in the massless limit. Since the form factor $B(t)$ is the gravitational analog of the Pauli form factor $F_2(t)$ and specifically corresponds to a helicity-flip transition, the absence of a helicity-flipping mechanism in the minimal coupling scheme leads to its vanishing. Consequently, any non-zero value for $B_N(t)$ would require the introduction of non-minimal coupling terms, such as higher-derivative Pauli-type interactions in the 5D action. The empirical observation that $B_N(t)$ is nearly zero suggests that the nucleon's interaction with gravity is dominated by these minimal, helicity-conserving holographic processes.

A closely related approach is light-front holographic QCD (LFHQCD), which establishes a mapping between the holographic description in 5D AdS space and light-front QCD \cite{deTeramond:2014asa, Dosch:2015nwa} (see Ref.~\cite{Brodsky:2014yha} for a review). Within this framework, the AGM $B_N(0)$ vanishes identically for each individual Fock sector, a result fundamentally ensured by the underlying light-front dynamics. The GFF $B_N(t)$ at finite $t$ also vanishes in this approach, as directly implied by the holographic correspondence. However, it is important to note that for theories other than QCD, the GFF $B(t)$ is not required to be zero \cite{Brodsky:2000ii, Burkert:2023wzr}. By performing a comparative analysis of the light-front wave functions (LFWFs) across different theoretical frameworks, we can gain deeper insight into the specific underlying dynamics that suppress $B_N(t)$ for the nucleon. This suggests that the smallness of this observable is not merely a general property of field theories, but a specific consequence of the nucleon's unique internal structure.

On the light-front side, the nucleon emerges as a bound state consisting of an active quark $q$ and a scalar diquark $D$. Within this representation, the GFF $B_N(t) = B_q(t) + B_D(t)$ is extracted from the spin-flip matrix element $T^{++}_{\uparrow\downarrow}$. The specific wave function representations for the quark and diquark contributions are given by \cite{Brodsky:2000ii, Kumar:2017dbf}:
\begin{equation}
\begin{split}
& \frac{q^1+iq^2}{2M}B_q(t) = \int \dd x \dd^2 \zeta_\perp \, (1-x) e^{i\sqrt{\frac{x}{1-x}}\vec q_\perp \cdot \vec\zeta_\perp}  \\
& \times \big[\psi_{\half}^{\downarrow*}(x, \vec\zeta_\perp)\psi_{\half}^\uparrow(x, \vec\zeta_\perp)
 + \psi_{-\half}^{\downarrow*}(x, \vec\zeta_\perp)\psi_{-\half}^\uparrow(x, \vec\zeta_\perp)\big] \\
& \frac{q^1+iq^2}{2M}B_D(t) = \int \dd x \dd^2 \zeta_\perp \, x e^{-i\sqrt{\frac{1-x}{x}}\vec q_\perp \cdot \vec\zeta_\perp}  \\
& \times \big[\psi_{\half}^{\downarrow*}(x, \vec\zeta_\perp)\psi_{\half}^\uparrow(x, \vec\zeta_\perp)
 + \psi_{-\half}^{\downarrow*}(x, \vec\zeta_\perp)\psi_{-\half}^\uparrow(x, \vec\zeta_\perp)\big].
\end{split}
\end{equation}
In these expressions, $t= -q^2_\perp$ and $x$ denotes the longitudinal momentum fraction of the active quark. The variable $\vec\zeta_\perp = \sqrt{x(1-x)}(\vec r_{D\perp} - \vec r_{q\perp})$ represents the transverse impact parameter, which defines the relative separation between the quark and the diquark. In the context of the AdS/QCD correspondence, this transverse variable is mapped directly to $z$ the fifth dimension of the AdS bulk. The LFWFs $\psi_{s_z}^{J_z}(x, \zeta_\perp)$ describe the nucleon state with total light-front helicity $J_z$, where $s_z = \pm 1/2$ is the helicity of the active quark. The system is characterized by two independent LFWFs, $\psi_+$ and $\psi_-$, corresponding to orbital angular momentum states $L_z = 0$ and $L_z = \pm 1$, respectively (where $L_z = J_z - s_z$). Crucially, one can verify that the total AGM vanishes at the forward limit, $B_N(0) = 0$, independent of the specific functional form of the LFWFs. This result is consistent with the general requirements of EP as realized in light-front dynamics.

 In LFHQCD, the wave functions adopt a remarkably simple form where the explicit dependence on the longitudinal momentum fraction $x$ effectively drops out \cite{Brodsky:2014yha}:
 \begin{equation}\label{eqn:holographic_lfwf}
\begin{split}
\psi_{+\half}^\uparrow(x, \vec\zeta_\perp) =\,& \psi_+(\zeta_\perp), \\
\psi_{-\half}^\uparrow(x, \vec\zeta_\perp) =\,& +i\psi_-(\zeta_\perp)e^{+i\theta_\zeta}, \\
\psi_{+\half}^\downarrow(x, \vec\zeta_\perp) =\,& -i\psi_-(\zeta_\perp)e^{-i\theta_\zeta}, \\
\psi_{-\half}^\downarrow(x, \vec\zeta_\perp) =\,& \psi_+(\zeta_\perp).
\end{split}
\end{equation}
Then the GFF $B(t)$ is expressed as an overlap integral of the orbital angular momentum components $\psi_+$ ($L_z=0$) and $\psi_-$ ($L_z=1$):
\begin{equation}\label{eqn:LFHQCD_B}
\begin{split}
B(t) = M\int \zeta_\perp \dd\zeta_\perp \mathcal B^{(0)}(\zeta_\perp, t) \psi_+(\zeta_\perp) \psi_-(\zeta_\perp).
\end{split}
\end{equation}
The kernel $\mathcal{B}^{(0)}(z, t)$ represents the ``bare" part of the holographic current, which consists of distinct quark and diquark contributions: $\mathcal{B}^{(0)}(z, -Q^2) = \mathcal{B}^{(0)}_q(z, -Q^2) + \mathcal{B}^{(0)}_D(z, -Q^2)$. These components are defined via the following integrals involving the Bessel function $J_1(z)$:
\begin{equation}
\begin{split}
\mathcal B^{(0)}_q(z, -Q^2) =\,& -\frac{8\pi}{Q} \int \dd x (1-x) J_1(\sqrt{\frac{x}{1-x}}zQ), \\
\mathcal B^{(0)}_D(z, -Q^2) =\,& \frac{8\pi}{Q} \int \dd x \, x J_1(\sqrt{\frac{1-x}{x}}zQ). \\
\end{split}
\end{equation}
Within the framework of light-front holography, $\mathcal{B}^{(0)}$ is identified as the valence contribution to a more general holographic current $\mathcal{B}(z, t)$, which in principle incorporates non-perturbative corrections beyond the $|qD\rangle$ Fock sector. Crucially, it can be analytically demonstrated that the sum of these quark and diquark kernels is zero:
\begin{equation}
 \mathcal B^{(0)}_q(z, -Q^2) +  \mathcal B^{(0)}_D(z, -Q^2) = 0.
\end{equation}
As a result, the total current $\mathcal{B}^{(0)}(z, -Q^2)$ vanishes identically. This leads to a vanishing GFF $B_N(t)$ for all finite $t$, providing a microscopic light-front derivation that is perfectly consistent with the predictions of holographic QCD in the 5D bulk \cite{Abidin:2009hr}.

A central assumption in the above derivation in LFHQCD is that the wave functions depend exclusively on the holographic variable $\zeta_\perp$. However, as demonstrated in Refs.~\cite{Li:2021jqb, deTeramond:2021yyi}, a longitudinal wave function $X(x)$ is generally required to account for longitudinal dynamics. Incorporating this component, the modified wave functions are expressed as:
\begin{equation}\label{eqn:holographic_lfwf}
\begin{split}
\psi_{+\half}^\uparrow(x, \vec\zeta_\perp) =\,& \psi_+(\zeta_\perp)X_+(x), \\
\psi_{-\half}^\uparrow(x, \vec\zeta_\perp) =\,& +i\psi_-(\zeta_\perp)e^{+i\theta_\zeta}X_-(x), \\
\psi_{+\half}^\downarrow(x, \vec\zeta_\perp) =\,& -i\psi_-(\zeta_\perp)e^{-i\theta_\zeta}X_-(x), \\
\psi_{-\half}^\downarrow(x, \vec\zeta_\perp) =\,& \psi_+(\zeta_\perp)X_+(x).
\end{split}
\end{equation}
The original holographic wave functions are recovered by setting $X_\pm(x) = 1$. With this generalization, the GFF $B_N(t)$ maintains its structure from Eq.~(\ref{eqn:LFHQCD_B}), but the holographic current is redefined as:
\begin{multline}\label{eqn:holographic_current}
\mathcal B^{(0)}(z, -Q^2)
= \frac{8\pi}{Q} \int \dd x (1-x) J_1(\sqrt{\frac{x}{1-x}}zQ) \\
\times \big[X_+(1-x)X_-(1-x) - X_+(x)X_-(x)\big].
\end{multline}
The antisymmetric factor $X_+(x)X_-(x) - X_+(1-x)X_-(1-x)$ acts as a primary suppression mechanism for the integral. This suppression can be understood through two specific limits. First, as $Q^2 \to 0$, the Bessel function term $(1-x)J_1(\sqrt{\frac{x}{1-x}}zQ)$ scales as $\frac{1}{2}\sqrt{x(1-x)}zQ$. In this limit, the $x$-integral vanishes regardless of the functional form of $X_\pm(x)$, yielding a vanishing AGM $B_N(0)$ in accordance with EP. Second, if $X_+(x)X_-(x)$ is symmetric around $x = 1/2$, the antisymmetric factor becomes zero, causing $B_N(t)$ to vanish identically for all $t$. Importantly, because both mechanisms operate independently of the precise shape of the wave functions, the observed suppression is a robust feature of the framework.

\begin{figure}
\centering
\includegraphics[width=0.45\textwidth]{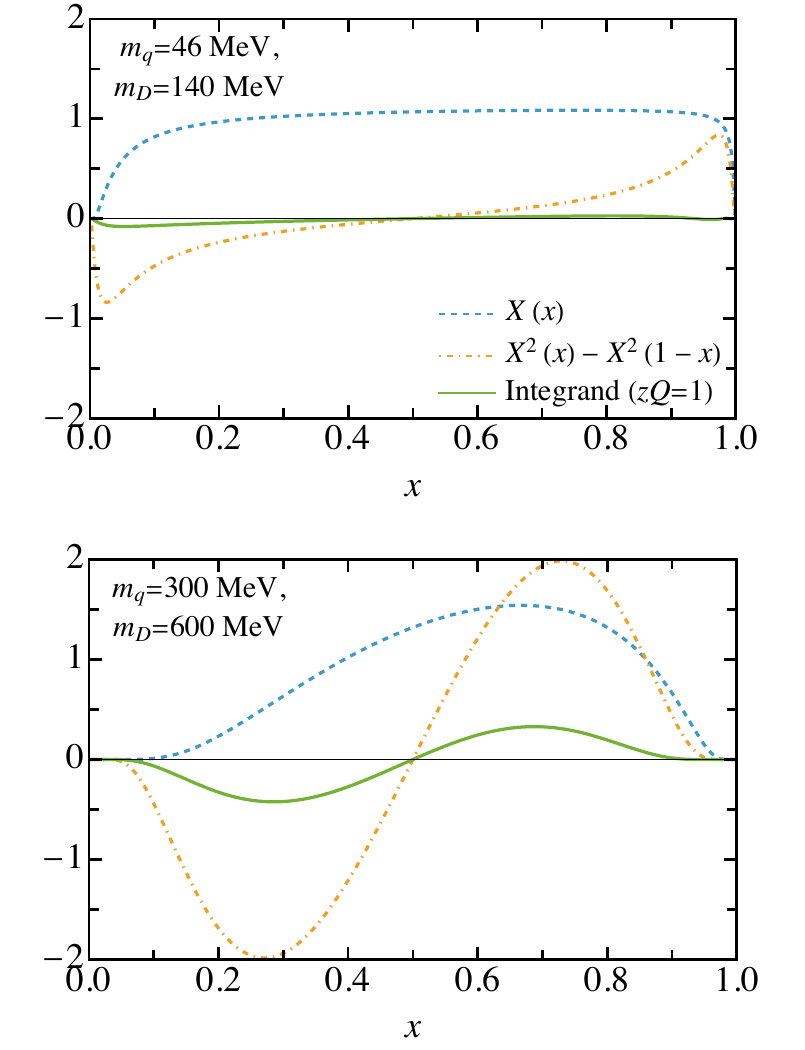}
\caption{The longitudinal wave function $X_\pm(x) = X(x)$, the corresponding suppression factor $X^2(x)-X^2(1-x)$, and the integrand of Eq.~(\ref{eqn:holographic_current}) at $zQ = 1$ with two sets of mass parameters: (\textit{Top}) $m_q = 46$ MeV and $m_D = 140$ MeV; (\textit{Bottom}) $m_q = 300$ MeV, $m_D = 600$ MeV. The former are the values adopted in LFHQCD and the latter are typical values adopted in constituent quark model. }
\label{fig:IMA}
\end{figure}

In LFHQCD, both the active quark and the diquark cluster are massless. Finite quark masses cause the deviation of the longitudinal wave function from unity, becoming asymmetric.
A standard ansatz for the longitudinal wave function in this framework is \cite{deTeramond:2021yyi}:
\begin{equation}\label{eqn:IMA}
X_\pm(x) = X(x) \equiv N \exp\Big\{ - \frac{m_q^2}{2\kappa^2 (1-x)} - \frac{m_D^2}{2\kappa^2 x} \Big\},
\end{equation}
where $N$ is a normalization constant, $\kappa = 0.54$ GeV represents the confining strength, and $m_q = 46$ MeV and $m_D = 140$ MeV are the quark and diquark masses, respectively. Despite the lack of perfect symmetry in this ansatz, the integral remains significantly suppressed due to the antisymmetric factor, as illustrated in Fig.~\ref{fig:IMA}. For comparison, Fig.~\ref{fig:IMA} also presents a second set of longitudinal wave functions utilizing masses typical of the constituent quark model ($m_q = 300$ MeV, $m_D = 600$ MeV). While these constituent-like masses increase the magnitude of the integrand, the total integral remains small because of internal cancellations within the antisymmetric weighting.
Figure~\ref{fig:GFF_B} illustrates the numerical results for the GFF $B_N(t)$, obtained by incorporating the longitudinal wave function from Eq.~(\ref{eqn:IMA}) into the LFHQCD framework. The result explicitly demonstrates the suppression of $B_N(t)$ across the physical range of momentum transfer.

\begin{figure}
\centering
\includegraphics[width=0.45\textwidth]{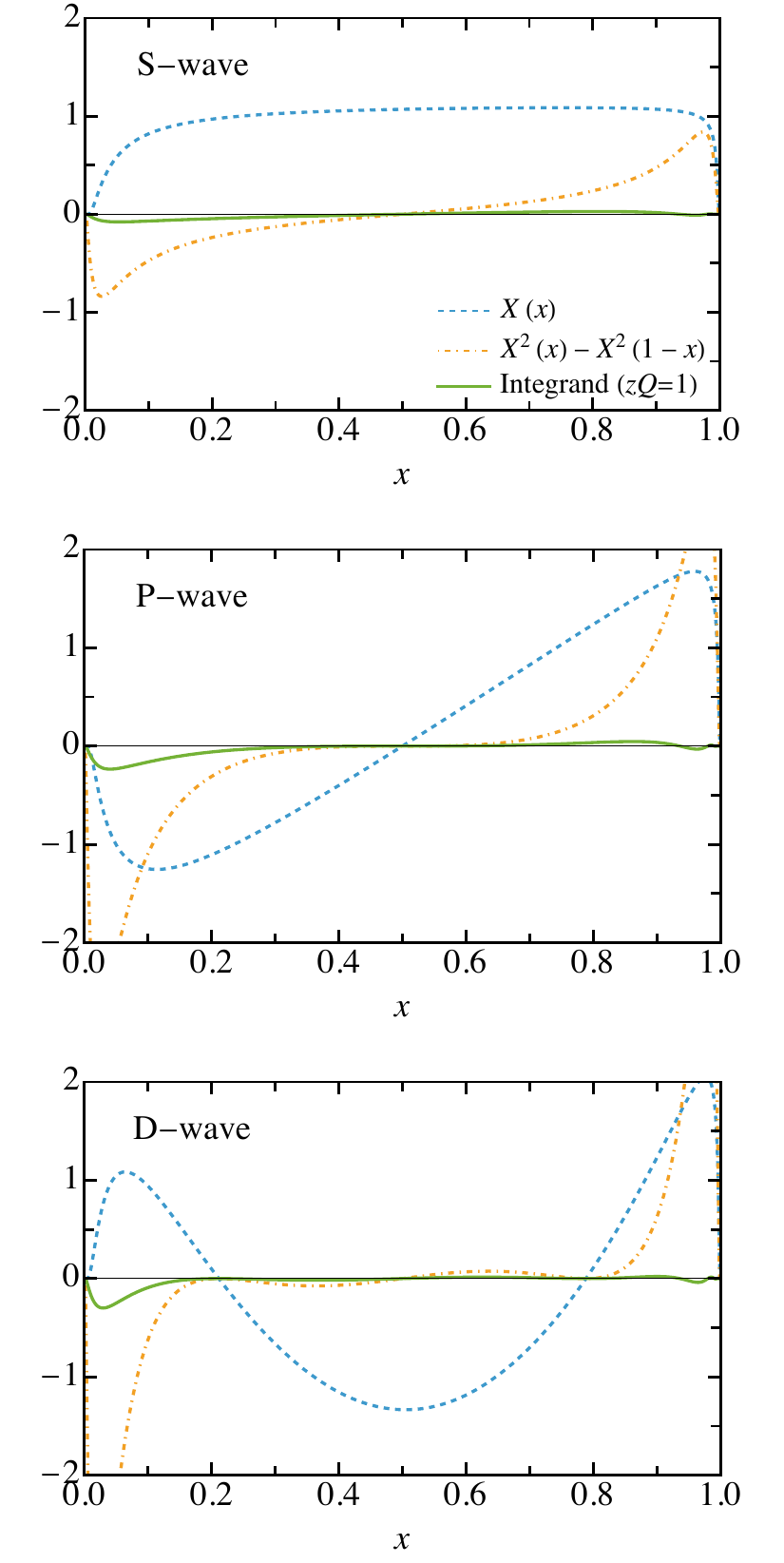}
\caption{Different partial waves of the longitudinal wave function $X_\pm(x) = X(x)$, and the corresponding suppression factor $X^2(x)-X^2(1-x)$ and the integrand of Eq.~(\ref{eqn:holographic_current}) at $zQ = 1$. From \textit{Top} to \textit{Bottom}, the partial waves are: S-wave ($n=0$), P-wave ($n=1$) and D-wave ($n=2$). The S-wave is given by Eq.~(\ref{eqn:IMA}), with $m_q = 46$ MeV and $m_D = 140$ MeV. The $n$-th partial wave is given by multiplying the Eq.~(\ref{eqn:IMA}) by the Legendre polynomial $P_n(2x-1)$ with a proper normalization. }
\label{fig:PWA}
\end{figure}

\begin{figure}
\centering
\includegraphics[width=0.45\textwidth]{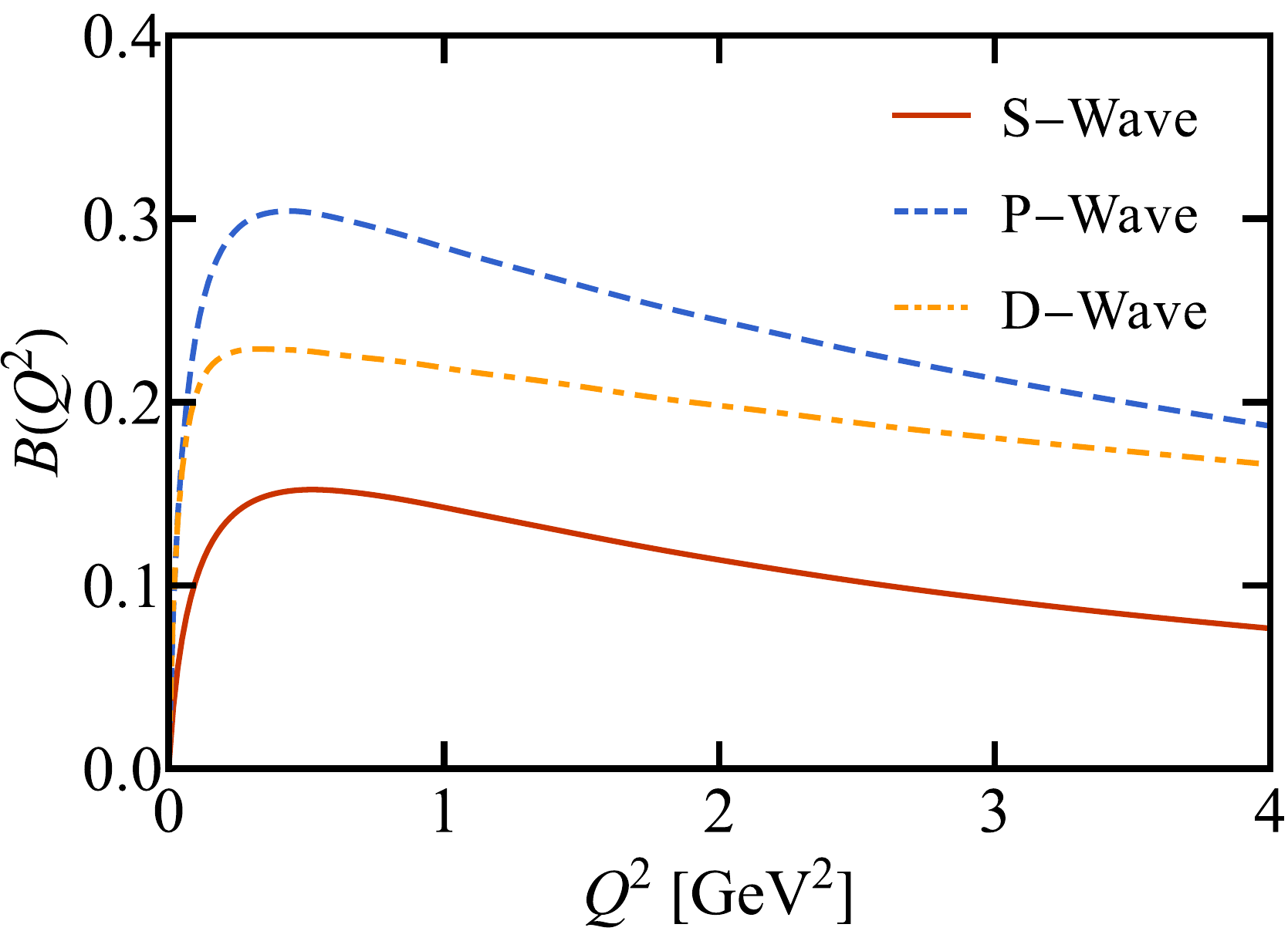}
\caption{Gravitational form factor $B(t)$ calculated from various partial wave contributions to the longitudinal wave function. These wave functions are derived using an invariant mass ansatz, where the $n$-th partial wave is represented by the Legendre polynomial $P_n(2x-1)$. }
\label{fig:GFF_PWA}
\end{figure}

In Fig.~\ref{fig:PWA}, we compare the integrands from different partial waves. They are obtained by multiplying the wave function $X(x)$ given in Eq.~(\ref{eqn:IMA}) by the Legendre polynomial $P_n(2x-1)$, where $n=0, 1, 2$ for S-, P-, D-waves, with a proper normalization.
As shown in Fig.~\ref{fig:PWA}, the longitudinal wave functions and corresponding integrands for S-, P-, and D-waves reveal that the S-wave component undergoes the most dramatic suppression. Consequently, within LFHQCD, the observed smallness of the GFF $B_N(t)$ at finite $t$ serves as a signature of the nucleon's dominant S-wave character. Figure~\ref{fig:GFF_PWA} compares the GFF $B(t)$ across various longitudinal partial waves. Notably, the S-wave contribution is significantly suppressed in comparison to the higher-order partial waves.

In this note, we discussed the smallness of the gravitational form factor $B_N(t)$ of the nucleon. This observable serves as a fundamental probe of the nucleon’s spin structure. While the equivalence principle strictly mandates that the total anomalous gravitomagnetic moment must vanish in the forward limit, $B_N(0) = 0$, the mechanisms responsible for its persistent smallness at finite momentum transfer have long remained a subject of theoretical interest. Throughout this analysis, we have demonstrated that this suppression is not accidental but is instead deeply rooted in the underlying structures of the nucleon.

In the context of GPDs, the smallness of $B_N(t)$ is largely attributed to a flavor cancellation between the $u$ and $d$ quark sectors. Meanwhile, in holographic QCD, the vanishing of $B_N(t)$ under minimal coupling emerges as a direct consequence of helicity conservation in the 5D bulk. However, neither interpretation is entirely satisfactory: the former relies on fine-tuning within the model parameters, while the latter is strictly limited to the minimal coupling scheme --- conflicting with lattice QCD results and other models where $B_N(t)$ is non-zero.

By bridging these descriptions through LFHQCD, we demonstrate that this suppression is analytically tied to an antisymmetric factor inherent in the longitudinal dynamics. At the forward limit ($t = 0$), the suppression is exact as required by the EP. At finite momentum transfer $t$, the integral governing $B_N(t)$ vanishes identically if the longitudinal wave functions are symmetric and remain heavily suppressed even when realistic mass scales are introduced.
This suggests that the near-zero value of $B_N(t)$ is a distinctive signature of the nucleon’s dominant S-wave character. While our analysis is grounded in LFHQCD, the underlying mechanism presented here is remarkably robust: it relies on fundamental symmetry properties and the relativistic structure of the nucleon rather than the specific fine-tuning of model parameters. Consequently, this mechanism likely remains applicable to a broader class of relativistic constituent models where $B_N(t)$ is small but non-vanishing.

Looking forward, these results provide a clear avenue for exploring the gravitational structure of more complex systems. A natural extension of this work involves the study of excited baryonic states, such as $N^*$ resonances, where significant P-wave and D-wave components are expected to generate much larger, non-vanishing GFF $B(t)$ signatures. Furthermore, the role of non-minimal gravitational couplings in the holographic action warrants a more detailed investigation to account for the subtle, non-zero values suggested by recent lattice QCD simulations. As new experimental data from the electron-ion colliders and Jefferson Lab begin to constrain the energy-momentum tensor through processes like near-threshold $J/\psi$ production, the light-front framework will be essential for interpreting the fundamental stress and spin distributions of hadronic matter.

\section*{Acknowledgements}
The authors thank C.D.~Roberts and Xionghui Cao for providing the gravitational form factor data calculated via Dyson-Schwinger equations (DSEs) and dispersion relations (DR), respectively.
This work is supported by the National Natural Science Foundation of China (NSFC) under Grant Nos. 12375081, 12247103, and by the Chinese Academy of Sciences under Grant No. YSBR-101.
X.-h.~Cao is supported by the NSFC under Grant No.~125B2111.
C.~M. is supported by new faculty start up funding the Institute of Modern Physics, Chinese Academy of Sciences, Grants No. E129952YR0.

\end{document}